\documentclass[nofootinbib,prl,superscriptaddress,a4paper,twocolumn]{revtex4-1}
\usepackage{geometry}
\usepackage[english]{babel}
\usepackage[latin1]{inputenc}
\usepackage{amsmath, amssymb, amsfonts, mathrsfs}
\usepackage{graphicx}

\usepackage[colorlinks,breaklinks]{hyperref}
\makeatletter
\newcommand\org@hypertarget{}
\let\org@hypertarget\hypertarget
\renewcommand\hypertarget[2]{%
  \Hy@raisedlink{\org@hypertarget{#1}{}}#2%
  }
\makeatother

\usepackage{mathtools}

\usepackage{cleveref}

\usepackage{algpseudocode}
\usepackage{algorithm}
\algrenewcommand\algorithmicrequire{\textbf{Assume:}}

\usepackage{braket}
\usepackage{float} 
\usepackage{silence}
\usepackage{url}

\usepackage{tabulary}
\usepackage{amsfonts}

\newcounter{algostep}
\newlength{\stepwidth}

\graphicspath{{images/}}
\usepackage{xcolor}
\usepackage{exscale}
\usepackage{latexsym}
\usepackage{bbm}
\usepackage{xspace}

\def\pmx{\begin{pmatrix}}
\def\emx{\end{pmatrix}}

\definecolor{darkgreen}{RGB}{0,130,0}

\begin{document} 

\title{Universal Quantum Computation by a Single Photon}

\author{Xiaoqin Gao}
\email{xiaoqin.gao@univie.ac.at}
\affiliation{Institute for Quantum Optics and Quantum Information (IQOQI), Austrian Academy of Sciences, Boltzmanngasse 3, 1090 Vienna, Austria.}
\affiliation{Vienna Center for Quantum Science \& Technology (VCQ), Faculty of Physics, University of Vienna, Boltzmanngasse 5, 1090 Vienna, Austria.}
\affiliation{National Mobile Communications Research Laboratory, Quantum Information Research Center, Southeast University, Sipailou 2, 210096 Nanjing, China.}

\author{Zhengwei Liu}
\email{liuzhengwei@mail.tsinghua.edu.cn}
\affiliation{YMSC and Department of Mathematics, Tsinghua University, Beijing 100084, China}

\begin{abstract}
We use one photon to simulate an $n$-qubit quantum system for the first time. We propose a new scheme to realize universal quantum computation in polynomial time $O(n^5)$. A generating set of gates can be realized with high accuracy in the lab. We conclude that photonic quantum computation is one of the promising approaches to universal quantum computation. 
\end{abstract}

\date{\today}
\maketitle

\section{Introduction}

Here we present a new experimental method for universal quantum computation represented by a single photon carrying orbital angular momentum (OAM). We use one photon to simulate a multi-body quantum system for the first time. We implement an $n$-qubit by the superposition of $2^n$ OAM modes of a single photon; a generating set of gates for universal quantum computation on $n$-qubits by elementary operations on a single photon. Experimentally these operations can be done with accessible optical components in the lab with high accuracy. Arbitrary $n$-qubit gate can be implemented by at most $O(n^5)$ elementary operations, therefore our scheme is efficient for universal quantum computation.

Quantum computation has been promised to increase greatly the efficiency of solving problems such as factoring large integers, combinatorial optimization and quantum physics simulation.
A very important problem is factoring large integers, and the best known classical randomized algorithms to find a factor of a composite number $d$ needs to run in time $2^{O({(\log d)}^{1\over 3})}$ \cite {lenstra1993development}. Remarkably, Shor introduced an algorithm using the quantum Fourier transform which factorizes integers only in time $O((\log d)^3)$ \cite{shor1994algorithms}. 

People have attempt different experimental approaches to implement  quantum Fourier transform and quantum computations in general.
In two-level quantum systems, any $n$-qubit universal quantum computation can be performed using a series of single qubit rotations and two qubits controlled-NOT (CNOT) gates \cite{barenco1995elementary}. 
It is particularly easy to observe quantum effects in optical systems, and photonic quantum computation has been proved to be performed using linear optics \cite {kok2007linear, knill2001scheme}. 
However, it has been a long standing open question that how to implement a universal quantum computation via multi-photons. A major problem is that the highest efficiency of implementing an accurate 2-qubits CNOT gate is $25\%$ due to the post selection \cite {o2003demonstration,  gasparoni2004realization}. Counting this probability, it essentially takes $O(d^2)$ times to implement an arbitrary $n$-qubit gate, which requires at least $n-1$ CNOT gates.

Recently, people have implemented an $n$-qubit by the superposition of $d=2^n$ on a single photon with OAM \cite{garcia2011universal}, which is a multi-dimensional degree of freedom \cite {allen1992orbital}. 
However, their construction of an $n$-qubit gate requires at least $O(2^n)$ operations, as each operation acts on at most two states.
A better approach shows that $d$-dimensional Pauli-$X$ and -$Z$ gates can be actually constructed by $O(\log d)$ operations \cite {gao2019arbitrary, wang2015quantum}. It is naturally to ask that whether we can implement the quantum Fourier transform in polynomial time $\log d$. If so, then we can realize Shor's algorithm via a single photon. Furthermore, whether we can do universal quantum computation via a single photon. In this paper, we answer both questions positively. We provide a new scheme to implement universal quantum computation using at most $O((\log d)^5)$ times by a single photon. 

\section{Main idea}
Our main idea is using one photon to simulate a multi-body quantum system. The $2^n$ orthogonal states $\ket{c_{1}c_{2}\ldots c_{n}}$ of an $n$-qubit quantum computational basis were encoded as the OAM state $\ket{m}$, such that the following binary identification holds:
\begin{align}\label{Equ: BI}
m=\sum_{k=1}^{n} c_{n-k} 2^{k-1}
\end{align}
where $0\leq m\leq d-1$ and $c_k\in \{0,1\}$.

The generating set of gates here for realizing universal quantum computation includes arbitrary phase gate ($e^{i\theta Z}$, $Z$ stands for Pauli-$Z$ gate) and the Hadamard gate on the first qubit, the  control-$Z$ gate on the first two qubits, and the cyclic permutation for all qubits. All these gates can be implemented using accessible optical components in the laboratory. We show that arbitrary $n$-qubit gate can be realized by the generating gates in time $O((\log d)^5)$. 
Taking the conjugate of the Hadamard gate on $e^{i\theta Z}$, we can obtain $e^{i\theta X}$ ($X$ means Pauli-$X$ gate).
Then any 1-qubit gate can be realized as a composition $e^{i\theta_1 Z} e^{i\theta_2 X}e^{i\theta_3 Z}$, according to the Euler's formula. 
Within the conjugation by cyclic permutation using at most $O(\log d)$ times, we then can implement arbitrary 1-qubit gate on any qubit.
The swap gate can be obtained on the first two qubits by using three control-$Z$ gates and six Hadamard gates. 
Then all swap gates can be implemented on any adjacent 2-qubits using $O(\log d)$ cyclic permutations and on any non-adjacent 2-qubits in time $O((\log d)^2)$. 
It is known that arbitrary $n$-qubit gate can be generated by arbitrary 1-qubit gates and the CNOT gates on any two qubits in time $O((\log d)^3)$ \cite{barenco1995elementary}.
Therefore, we can realize universal quantum computation using at most $O((\log d)^5)$ times on a single photon.

\section{Generating set of gates}
Under the binary identification \eqref{Equ: BI}, we consider the superposition of OAM modes $\{\ket{0}, \ket{1}, \ket{2}, \cdots, \ket{d-1}\}$ of a single photon as an $n$-qubit.
We then identity the generating set of gates on $n$-qubits as operations on $d$ OAM modes.

\begin{figure}[!htbp]
\includegraphics[width=\columnwidth]{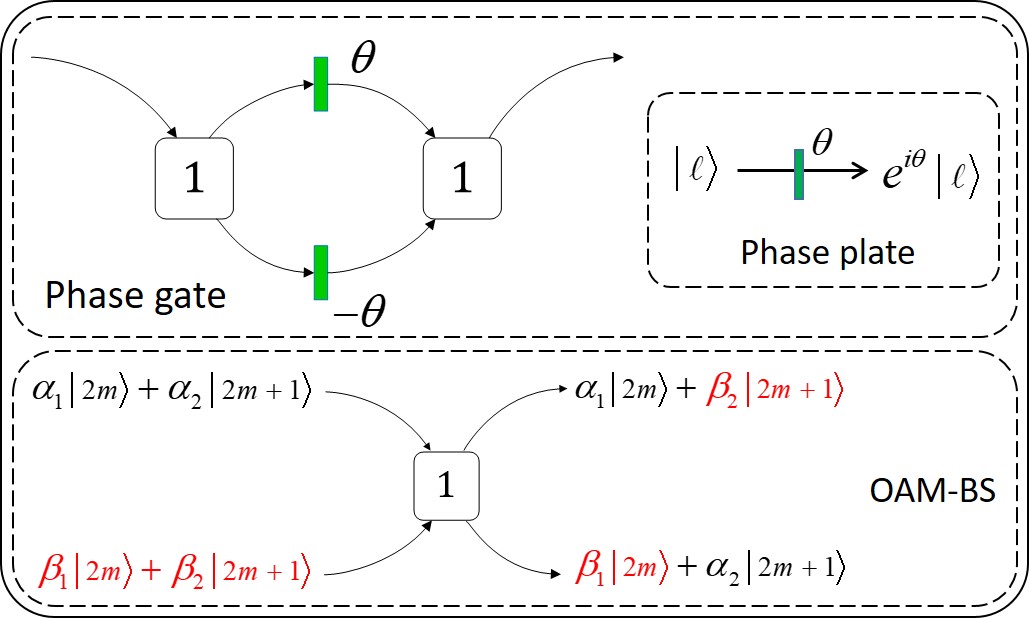}
 \vspace{-0.4cm}%
\caption{Arbitrary phase gate. Phase plate is an OAM-independent plate and the OAM-BS works as a parity sorter \cite{Leach2002}.}  
\centering
\label{fig1:figure1}
\vspace{-0.2cm}%
\end{figure}

Arbitrary phase gate $e^{i\theta Z}$ on the first qubit can be implemented experimentally as shown in figure \ref{fig1:figure1}. According to the binary identification \eqref{Equ: BI}, the phase gate $\rho_{\theta}$=$e^{i\theta Z} \otimes \mathbb{I} \otimes \mathbb{I} \cdots \otimes \mathbb{I}$ ($\mathbb{I}$ is an identity operation) will do the operation $\tilde{\rho}_{\theta}$ to the OAM states as follows:
\begin{equation}\label{phase gate}
\tilde{\rho}_{\theta}\ket{l}= 
\begin{cases}
   e^{i\theta}\ket{l}, & l \in even\\
   e^{-i\theta}\ket{l},              & l \in odd
\end{cases}
\end{equation}

\begin{figure}[!htbp]
\includegraphics[width=\columnwidth]{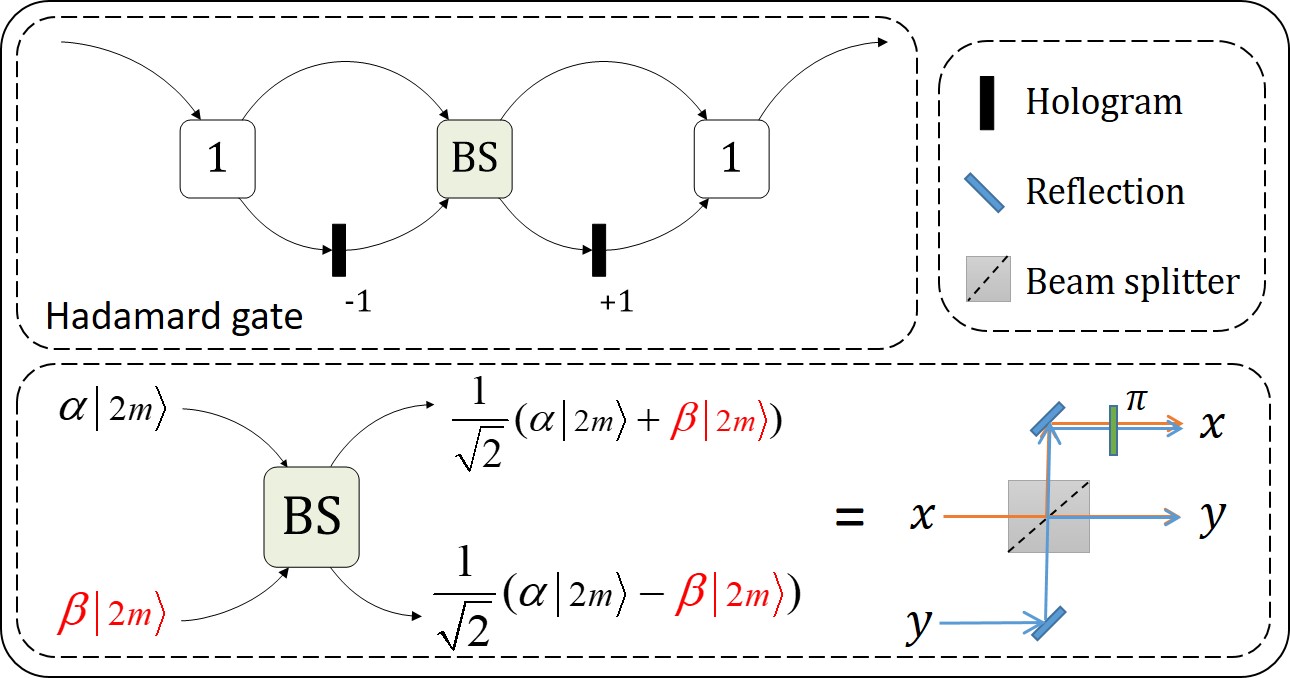}
 \vspace{-0.4cm}%
\caption{The hadamard gate works on the OAM of a single photon. The BS in green is an OAM-independent beam splitter, where $x$ and $y$ stand for paths.}  
\centering
\label{fig2:figure2}
\vspace{-0.2cm}%
\end{figure}

The hadamard ($H$) gate on the first qubit can be implemented experimentally by OAM-BS and OAM-independent beam splitter, as shown in figure \ref{fig2:figure2}. According to the binary identification \eqref{Equ: BI}, the Hadamard gate $\Lambda$=$H \otimes \mathbb{I} \otimes \mathbb{I} \cdots \otimes \mathbb{I}$ will do the operation $\tilde{\Lambda}$ for the OAM states as follows:
\begin{equation}\label{Hadamard gate}
\begin{split}
\tilde{\Lambda}\ket{2m} &=  \frac{1}{\sqrt{2}} (\ket{2m} + \ket{2m+1})\\
\tilde{\Lambda}\ket{2m+1} &=  \frac{1}{\sqrt{2}} (\ket{2m} - \ket{2m+1})
\end{split}
\end{equation}
where $0\leq m\leq {d\over2}-1$.

\begin{figure}[!htbp]
\includegraphics[width=\columnwidth]{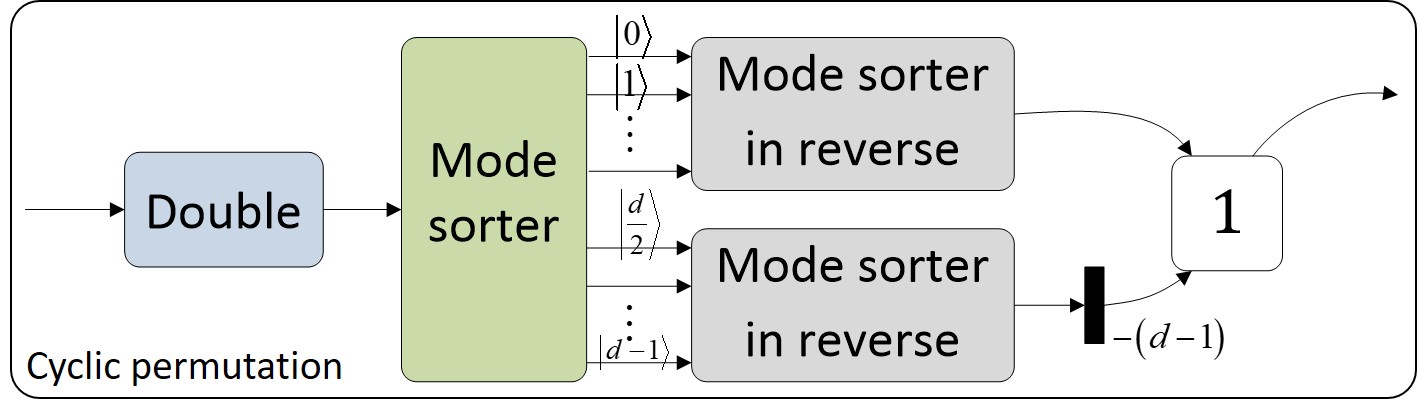}
 \vspace{-0.4cm}%
\caption{The cyclic permutation. Double transformation is used to expand the values of OAM modes to two times \cite{potovcek2015quantum}. Mode sorter is used for splitting all modes into each path \cite{berkhout2010efficient, mirhosseini2013efficient}. Then all modes can be combined via a mode sorter in reverse \cite{fickler2014interface}.}  
\centering
\label{fig3:figure3}
\vspace{-0.2cm}%
\end{figure}

The cyclic permutation on $n$-qubits can be implemented experimentally as shown in figure \ref{fig3:figure3}. According to the binary identification \eqref{Equ: BI}, the cyclic permutation ${\Gamma}\ket{c_{1}c_{2}\ldots c_{n}}$=$\ket{c_{1}c_{3}\ldots c_{n-1}c_{2}c_{4}\ldots c_{n}}$ will operate the OAM states as an operation $\tilde{\Gamma}$ as follows:
\begin{equation}\label{Cyclic permutation}
\tilde{\Gamma}\ket{l}= 
\begin{cases}
   \ket{2l}, & 0\leq l \leq {{d\over2} - 1}\\
   \ket{2l-d+1},              & {{d\over2} - 1}< l \leq {d - 1}
\end{cases}
\end{equation}

\begin{figure}[!htbp]
\includegraphics[width=0.4\textwidth]{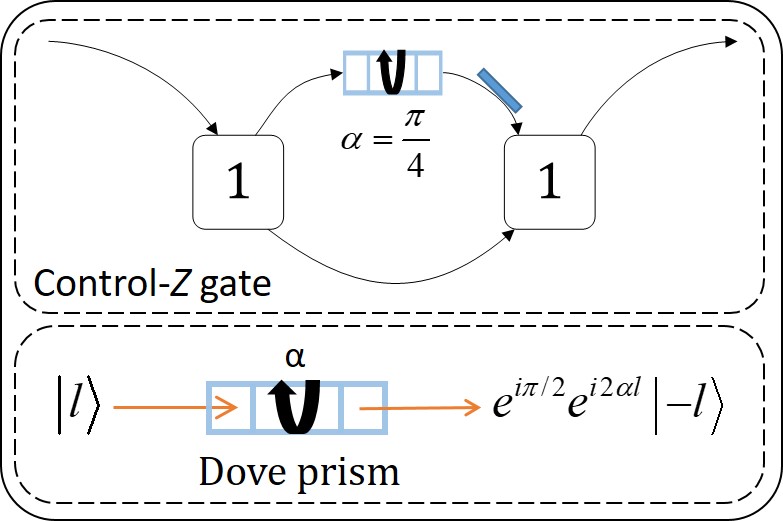}
 \vspace{0.1cm}%
\caption{The control-$Z$ gate. Dove Prism is a mode-based phase operation \cite{gonzalez2006dove}.}  
\centering
\label{fig4:figure4}
\vspace{-0.2cm}%
\end{figure}

The control-$Z$ ($CZ$) gate (-1, 1, 1, 1) on the first two qubits can be implemented experimentally as shown in figure \ref{fig4:figure4}. According to the binary identification \eqref{Equ: BI}, the $CZ$ gate $\Omega$=$CZ \otimes \mathbb{I} \otimes \mathbb{I} \cdots \otimes \mathbb{I}$ will do the corresponding operation $\tilde{\Omega}$ on the OAM states as follows:
\begin{equation}\label{CZ gate}
\begin{split}
\tilde{\Omega}\ket{4m} &= -\ket{4m}\\
\tilde{\Omega}\ket{4m+j} &=\ket{4m+j} 
\end{split}
\end{equation}
where $0\leq m\leq {d\over4}-1$ and $j=1,2,3$.

Using these generating set of gates, we can realize universal quantum computation on the OAM of a single photon in time at most $O((\log d)^5)$ as we discussed in the main idea.

\section{Conclusion}

We introduce a new scheme to simulate a multi-qubit system by a single photon.  We implement universe quantum computation on the OAM of a single photon in polynomial time $\log d$ for the first time. All operations on the single photon can be done with accessible optical components in the lab. Our scheme is very promising for realizing universal photonic quantum computation. 

\section{ACKNOWLEDGMENTS}

The authors thank Anton Zeilinger,  Borivoje Daki\'c, Arthur Jaffe, Hui Wang, Mehul Malik and Jaroslav Kysela for helpful discussions. XQ. G. thanks Bin Sheng and Zaichen Zhang for support. XQ. G. acknowledges support from the National Natural Science Foundation of China (Nos. 61571105, 61501109, and 61601119), the Scientific Research Foundation of Graduate School of Southeast University (No. YBJJ1710), and China Scholarship Council (CSC). ZW. L. was supported by Templeton Religion Trust under the grant TRT 159.

\bibliographystyle{apsrev4-1}
\bibliography{refs}

\begin{thebibliography}{17}%
\makeatletter
\providecommand \@ifxundefined [1]{%
 \@ifx{#1\undefined}
}%
\providecommand \@ifnum [1]{%
 \ifnum #1\expandafter \@firstoftwo
 \else \expandafter \@secondoftwo
 \fi
}%
\providecommand \@ifx [1]{%
 \ifx #1\expandafter \@firstoftwo
 \else \expandafter \@secondoftwo
 \fi
}%
\providecommand \natexlab [1]{#1}%
\providecommand \enquote  [1]{``#1''}%
\providecommand \bibnamefont  [1]{#1}%
\providecommand \bibfnamefont [1]{#1}%
\providecommand \citenamefont [1]{#1}%
\providecommand \href@noop [0]{\@secondoftwo}%
\providecommand \href [0]{\begingroup \@sanitize@url \@href}%
\providecommand \@href[1]{\@@startlink{#1}\@@href}%
\providecommand \@@href[1]{\endgroup#1\@@endlink}%
\providecommand \@sanitize@url [0]{\catcode `\\12\catcode `\$12\catcode
  `\&12\catcode `\#12\catcode `\^12\catcode `\_12\catcode `\%12\relax}%
\providecommand \@@startlink[1]{}%
\providecommand \@@endlink[0]{}%
\providecommand \url  [0]{\begingroup\@sanitize@url \@url }%
\providecommand \@url [1]{\endgroup\@href {#1}{\urlprefix }}%
\providecommand \urlprefix  [0]{URL }%
\providecommand \Eprint [0]{\href }%
\providecommand \doibase [0]{http://dx.doi.org/}%
\providecommand \selectlanguage [0]{\@gobble}%
\providecommand \bibinfo  [0]{\@secondoftwo}%
\providecommand \bibfield  [0]{\@secondoftwo}%
\providecommand \translation [1]{[#1]}%
\providecommand \BibitemOpen [0]{}%
\providecommand \bibitemStop [0]{}%
\providecommand \bibitemNoStop [0]{.\EOS\space}%
\providecommand \EOS [0]{\spacefactor3000\relax}%
\providecommand \BibitemShut  [1]{\csname bibitem#1\endcsname}%
\let\auto@bib@innerbib\@empty
\bibitem [{\citenamefont {Lenstra}\ \emph {et~al.}(1993)\citenamefont
  {Lenstra}, \citenamefont {Hendrik~Jr} \emph
  {et~al.}}]{lenstra1993development}%
  \BibitemOpen
  \bibfield  {author} {\bibinfo {author} {\bibfnamefont {A.~K.}\ \bibnamefont
  {Lenstra}}, \bibinfo {author} {\bibfnamefont {W.}~\bibnamefont {Hendrik~Jr}},
   \emph {et~al.},\ }\href@noop {} {\emph {\bibinfo {title} {The development of
  the number field sieve}}},\ Vol.\ \bibinfo {volume} {1554}\ (\bibinfo
  {publisher} {Springer Science \& Business Media},\ \bibinfo {year}
  {1993})\BibitemShut {NoStop}%
\bibitem [{\citenamefont {Shor}(1994)}]{shor1994algorithms}%
  \BibitemOpen
  \bibfield  {author} {\bibinfo {author} {\bibfnamefont {P.~W.}\ \bibnamefont
  {Shor}},\ }in\ \href@noop {} {\emph {\bibinfo {booktitle} {Proceedings 35th
  annual symposium on foundations of computer science}}}\ (\bibinfo
  {organization} {Ieee},\ \bibinfo {year} {1994})\ pp.\ \bibinfo {pages}
  {124--134}\BibitemShut {NoStop}%
\bibitem [{\citenamefont {Barenco}\ \emph {et~al.}(1995)\citenamefont
  {Barenco}, \citenamefont {Bennett}, \citenamefont {Cleve}, \citenamefont
  {DiVincenzo}, \citenamefont {Margolus}, \citenamefont {Shor}, \citenamefont
  {Sleator}, \citenamefont {Smolin},\ and\ \citenamefont
  {Weinfurter}}]{barenco1995elementary}%
  \BibitemOpen
  \bibfield  {author} {\bibinfo {author} {\bibfnamefont {A.}~\bibnamefont
  {Barenco}}, \bibinfo {author} {\bibfnamefont {C.~H.}\ \bibnamefont
  {Bennett}}, \bibinfo {author} {\bibfnamefont {R.}~\bibnamefont {Cleve}},
  \bibinfo {author} {\bibfnamefont {D.~P.}\ \bibnamefont {DiVincenzo}},
  \bibinfo {author} {\bibfnamefont {N.}~\bibnamefont {Margolus}}, \bibinfo
  {author} {\bibfnamefont {P.}~\bibnamefont {Shor}}, \bibinfo {author}
  {\bibfnamefont {T.}~\bibnamefont {Sleator}}, \bibinfo {author} {\bibfnamefont
  {J.~A.}\ \bibnamefont {Smolin}}, \ and\ \bibinfo {author} {\bibfnamefont
  {H.}~\bibnamefont {Weinfurter}},\ }\href@noop {} {\bibfield  {journal}
  {\bibinfo  {journal} {Physical review A}\ }\textbf {\bibinfo {volume} {52}},\
  \bibinfo {pages} {3457} (\bibinfo {year} {1995})}\BibitemShut {NoStop}%
\bibitem [{\citenamefont {Kok}\ \emph {et~al.}(2007)\citenamefont {Kok},
  \citenamefont {Munro}, \citenamefont {Nemoto}, \citenamefont {Ralph},
  \citenamefont {Dowling},\ and\ \citenamefont {Milburn}}]{kok2007linear}%
  \BibitemOpen
  \bibfield  {author} {\bibinfo {author} {\bibfnamefont {P.}~\bibnamefont
  {Kok}}, \bibinfo {author} {\bibfnamefont {W.~J.}\ \bibnamefont {Munro}},
  \bibinfo {author} {\bibfnamefont {K.}~\bibnamefont {Nemoto}}, \bibinfo
  {author} {\bibfnamefont {T.~C.}\ \bibnamefont {Ralph}}, \bibinfo {author}
  {\bibfnamefont {J.~P.}\ \bibnamefont {Dowling}}, \ and\ \bibinfo {author}
  {\bibfnamefont {G.~J.}\ \bibnamefont {Milburn}},\ }\href@noop {} {\bibfield
  {journal} {\bibinfo  {journal} {Reviews of Modern Physics}\ }\textbf
  {\bibinfo {volume} {79}},\ \bibinfo {pages} {135} (\bibinfo {year}
  {2007})}\BibitemShut {NoStop}%
\bibitem [{\citenamefont {Knill}\ \emph {et~al.}(2001)\citenamefont {Knill},
  \citenamefont {Laflamme},\ and\ \citenamefont {Milburn}}]{knill2001scheme}%
  \BibitemOpen
  \bibfield  {author} {\bibinfo {author} {\bibfnamefont {E.}~\bibnamefont
  {Knill}}, \bibinfo {author} {\bibfnamefont {R.}~\bibnamefont {Laflamme}}, \
  and\ \bibinfo {author} {\bibfnamefont {G.~J.}\ \bibnamefont {Milburn}},\
  }\href@noop {} {\bibfield  {journal} {\bibinfo  {journal} {nature}\ }\textbf
  {\bibinfo {volume} {409}},\ \bibinfo {pages} {46} (\bibinfo {year}
  {2001})}\BibitemShut {NoStop}%
\bibitem [{\citenamefont {O'Brien}\ \emph {et~al.}(2003)\citenamefont
  {O'Brien}, \citenamefont {Pryde}, \citenamefont {White}, \citenamefont
  {Ralph},\ and\ \citenamefont {Branning}}]{o2003demonstration}%
  \BibitemOpen
  \bibfield  {author} {\bibinfo {author} {\bibfnamefont {J.~L.}\ \bibnamefont
  {O'Brien}}, \bibinfo {author} {\bibfnamefont {G.~J.}\ \bibnamefont {Pryde}},
  \bibinfo {author} {\bibfnamefont {A.~G.}\ \bibnamefont {White}}, \bibinfo
  {author} {\bibfnamefont {T.~C.}\ \bibnamefont {Ralph}}, \ and\ \bibinfo
  {author} {\bibfnamefont {D.}~\bibnamefont {Branning}},\ }\href@noop {}
  {\bibfield  {journal} {\bibinfo  {journal} {Nature}\ }\textbf {\bibinfo
  {volume} {426}},\ \bibinfo {pages} {264} (\bibinfo {year}
  {2003})}\BibitemShut {NoStop}%
\bibitem [{\citenamefont {Gasparoni}\ \emph {et~al.}(2004)\citenamefont
  {Gasparoni}, \citenamefont {Pan}, \citenamefont {Walther}, \citenamefont
  {Rudolph},\ and\ \citenamefont {Zeilinger}}]{gasparoni2004realization}%
  \BibitemOpen
  \bibfield  {author} {\bibinfo {author} {\bibfnamefont {S.}~\bibnamefont
  {Gasparoni}}, \bibinfo {author} {\bibfnamefont {J.-W.}\ \bibnamefont {Pan}},
  \bibinfo {author} {\bibfnamefont {P.}~\bibnamefont {Walther}}, \bibinfo
  {author} {\bibfnamefont {T.}~\bibnamefont {Rudolph}}, \ and\ \bibinfo
  {author} {\bibfnamefont {A.}~\bibnamefont {Zeilinger}},\ }\href@noop {}
  {\bibfield  {journal} {\bibinfo  {journal} {Physical review letters}\
  }\textbf {\bibinfo {volume} {93}},\ \bibinfo {pages} {020504} (\bibinfo
  {year} {2004})}\BibitemShut {NoStop}%
\bibitem [{\citenamefont {Garc{\'\i}a-Escart{\'\i}n}\ and\ \citenamefont
  {Chamorro-Posada}(2011)}]{garcia2011universal}%
  \BibitemOpen
  \bibfield  {author} {\bibinfo {author} {\bibfnamefont {J.~C.}\ \bibnamefont
  {Garc{\'\i}a-Escart{\'\i}n}}\ and\ \bibinfo {author} {\bibfnamefont
  {P.}~\bibnamefont {Chamorro-Posada}},\ }\href@noop {} {\bibfield  {journal}
  {\bibinfo  {journal} {Journal of Optics}\ }\textbf {\bibinfo {volume} {13}},\
  \bibinfo {pages} {064022} (\bibinfo {year} {2011})}\BibitemShut {NoStop}%
\bibitem [{\citenamefont {Allen}\ \emph {et~al.}(1992)\citenamefont {Allen},
  \citenamefont {Beijersbergen}, \citenamefont {Spreeuw},\ and\ \citenamefont
  {Woerdman}}]{allen1992orbital}%
  \BibitemOpen
  \bibfield  {author} {\bibinfo {author} {\bibfnamefont {L.}~\bibnamefont
  {Allen}}, \bibinfo {author} {\bibfnamefont {M.~W.}\ \bibnamefont
  {Beijersbergen}}, \bibinfo {author} {\bibfnamefont {R.}~\bibnamefont
  {Spreeuw}}, \ and\ \bibinfo {author} {\bibfnamefont {J.}~\bibnamefont
  {Woerdman}},\ }\href@noop {} {\bibfield  {journal} {\bibinfo  {journal}
  {Physical Review A}\ }\textbf {\bibinfo {volume} {45}},\ \bibinfo {pages}
  {8185} (\bibinfo {year} {1992})}\BibitemShut {NoStop}%
\bibitem [{\citenamefont {Gao}\ \emph {et~al.}(2019)\citenamefont {Gao},
  \citenamefont {Krenn}, \citenamefont {Kysela},\ and\ \citenamefont
  {Zeilinger}}]{gao2019arbitrary}%
  \BibitemOpen
  \bibfield  {author} {\bibinfo {author} {\bibfnamefont {X.}~\bibnamefont
  {Gao}}, \bibinfo {author} {\bibfnamefont {M.}~\bibnamefont {Krenn}}, \bibinfo
  {author} {\bibfnamefont {J.}~\bibnamefont {Kysela}}, \ and\ \bibinfo {author}
  {\bibfnamefont {A.}~\bibnamefont {Zeilinger}},\ }\href@noop {} {\bibfield
  {journal} {\bibinfo  {journal} {Physical Review A}\ }\textbf {\bibinfo
  {volume} {99}},\ \bibinfo {pages} {023825} (\bibinfo {year}
  {2019})}\BibitemShut {NoStop}%
\bibitem [{\citenamefont {Wang}\ \emph {et~al.}(2015)\citenamefont {Wang},
  \citenamefont {Cai}, \citenamefont {Su}, \citenamefont {Chen}, \citenamefont
  {Wu}, \citenamefont {Li}, \citenamefont {Liu}, \citenamefont {Lu},\ and\
  \citenamefont {Pan}}]{wang2015quantum}%
  \BibitemOpen
  \bibfield  {author} {\bibinfo {author} {\bibfnamefont {X.-L.}\ \bibnamefont
  {Wang}}, \bibinfo {author} {\bibfnamefont {X.-D.}\ \bibnamefont {Cai}},
  \bibinfo {author} {\bibfnamefont {Z.-E.}\ \bibnamefont {Su}}, \bibinfo
  {author} {\bibfnamefont {M.-C.}\ \bibnamefont {Chen}}, \bibinfo {author}
  {\bibfnamefont {D.}~\bibnamefont {Wu}}, \bibinfo {author} {\bibfnamefont
  {L.}~\bibnamefont {Li}}, \bibinfo {author} {\bibfnamefont {N.-L.}\
  \bibnamefont {Liu}}, \bibinfo {author} {\bibfnamefont {C.-Y.}\ \bibnamefont
  {Lu}}, \ and\ \bibinfo {author} {\bibfnamefont {J.-W.}\ \bibnamefont {Pan}},\
  }\href@noop {} {\bibfield  {journal} {\bibinfo  {journal} {Nature}\ }\textbf
  {\bibinfo {volume} {518}},\ \bibinfo {pages} {516} (\bibinfo {year}
  {2015})}\BibitemShut {NoStop}%
\bibitem [{\citenamefont {Leach}\ \emph {et~al.}(2002)\citenamefont {Leach},
  \citenamefont {Padgett}, \citenamefont {Barnett}, \citenamefont
  {Franke-Arnold},\ and\ \citenamefont {Courtial}}]{Leach2002}%
  \BibitemOpen
  \bibfield  {author} {\bibinfo {author} {\bibfnamefont {J.}~\bibnamefont
  {Leach}}, \bibinfo {author} {\bibfnamefont {M.~J.}\ \bibnamefont {Padgett}},
  \bibinfo {author} {\bibfnamefont {S.~M.}\ \bibnamefont {Barnett}}, \bibinfo
  {author} {\bibfnamefont {S.}~\bibnamefont {Franke-Arnold}}, \ and\ \bibinfo
  {author} {\bibfnamefont {J.}~\bibnamefont {Courtial}},\ }\href@noop {}
  {\bibfield  {journal} {\bibinfo  {journal} {Physical Review Letters}\
  }\textbf {\bibinfo {volume} {88}},\ \bibinfo {pages} {013601} (\bibinfo
  {year} {2002})}\BibitemShut {NoStop}%
\bibitem [{\citenamefont {Poto{\v{c}}ek}\ \emph {et~al.}(2015)\citenamefont
  {Poto{\v{c}}ek}, \citenamefont {Miatto}, \citenamefont {Mirhosseini},
  \citenamefont {Maga{\~n}a-Loaiza}, \citenamefont {Liapis}, \citenamefont
  {Oi}, \citenamefont {Boyd},\ and\ \citenamefont
  {Jeffers}}]{potovcek2015quantum}%
  \BibitemOpen
  \bibfield  {author} {\bibinfo {author} {\bibfnamefont {V.}~\bibnamefont
  {Poto{\v{c}}ek}}, \bibinfo {author} {\bibfnamefont {F.~M.}\ \bibnamefont
  {Miatto}}, \bibinfo {author} {\bibfnamefont {M.}~\bibnamefont {Mirhosseini}},
  \bibinfo {author} {\bibfnamefont {O.~S.}\ \bibnamefont {Maga{\~n}a-Loaiza}},
  \bibinfo {author} {\bibfnamefont {A.~C.}\ \bibnamefont {Liapis}}, \bibinfo
  {author} {\bibfnamefont {D.~K.}\ \bibnamefont {Oi}}, \bibinfo {author}
  {\bibfnamefont {R.~W.}\ \bibnamefont {Boyd}}, \ and\ \bibinfo {author}
  {\bibfnamefont {J.}~\bibnamefont {Jeffers}},\ }\href@noop {} {\bibfield
  {journal} {\bibinfo  {journal} {Physical review letters}\ }\textbf {\bibinfo
  {volume} {115}},\ \bibinfo {pages} {160505} (\bibinfo {year}
  {2015})}\BibitemShut {NoStop}%
\bibitem [{\citenamefont {Berkhout}\ \emph {et~al.}(2010)\citenamefont
  {Berkhout}, \citenamefont {Lavery}, \citenamefont {Courtial}, \citenamefont
  {Beijersbergen},\ and\ \citenamefont {Padgett}}]{berkhout2010efficient}%
  \BibitemOpen
  \bibfield  {author} {\bibinfo {author} {\bibfnamefont {G.~C.}\ \bibnamefont
  {Berkhout}}, \bibinfo {author} {\bibfnamefont {M.~P.}\ \bibnamefont
  {Lavery}}, \bibinfo {author} {\bibfnamefont {J.}~\bibnamefont {Courtial}},
  \bibinfo {author} {\bibfnamefont {M.~W.}\ \bibnamefont {Beijersbergen}}, \
  and\ \bibinfo {author} {\bibfnamefont {M.~J.}\ \bibnamefont {Padgett}},\
  }\href@noop {} {\bibfield  {journal} {\bibinfo  {journal} {Physical review
  letters}\ }\textbf {\bibinfo {volume} {105}},\ \bibinfo {pages} {153601}
  (\bibinfo {year} {2010})}\BibitemShut {NoStop}%
\bibitem [{\citenamefont {Mirhosseini}\ \emph {et~al.}(2013)\citenamefont
  {Mirhosseini}, \citenamefont {Malik}, \citenamefont {Shi},\ and\
  \citenamefont {Boyd}}]{mirhosseini2013efficient}%
  \BibitemOpen
  \bibfield  {author} {\bibinfo {author} {\bibfnamefont {M.}~\bibnamefont
  {Mirhosseini}}, \bibinfo {author} {\bibfnamefont {M.}~\bibnamefont {Malik}},
  \bibinfo {author} {\bibfnamefont {Z.}~\bibnamefont {Shi}}, \ and\ \bibinfo
  {author} {\bibfnamefont {R.~W.}\ \bibnamefont {Boyd}},\ }\href@noop {}
  {\bibfield  {journal} {\bibinfo  {journal} {Nature communications}\ }\textbf
  {\bibinfo {volume} {4}},\ \bibinfo {pages} {2781} (\bibinfo {year}
  {2013})}\BibitemShut {NoStop}%
\bibitem [{\citenamefont {Fickler}\ \emph {et~al.}(2014)\citenamefont
  {Fickler}, \citenamefont {Lapkiewicz}, \citenamefont {Huber}, \citenamefont
  {Lavery}, \citenamefont {Padgett},\ and\ \citenamefont
  {Zeilinger}}]{fickler2014interface}%
  \BibitemOpen
  \bibfield  {author} {\bibinfo {author} {\bibfnamefont {R.}~\bibnamefont
  {Fickler}}, \bibinfo {author} {\bibfnamefont {R.}~\bibnamefont {Lapkiewicz}},
  \bibinfo {author} {\bibfnamefont {M.}~\bibnamefont {Huber}}, \bibinfo
  {author} {\bibfnamefont {M.~P.}\ \bibnamefont {Lavery}}, \bibinfo {author}
  {\bibfnamefont {M.~J.}\ \bibnamefont {Padgett}}, \ and\ \bibinfo {author}
  {\bibfnamefont {A.}~\bibnamefont {Zeilinger}},\ }\href@noop {} {\bibfield
  {journal} {\bibinfo  {journal} {Nature communications}\ }\textbf {\bibinfo
  {volume} {5}},\ \bibinfo {pages} {4502} (\bibinfo {year} {2014})}\BibitemShut
  {NoStop}%
\bibitem [{\citenamefont {Gonz{\'a}lez}\ \emph {et~al.}(2006)\citenamefont
  {Gonz{\'a}lez}, \citenamefont {Molina-Terriza},\ and\ \citenamefont
  {Torres}}]{gonzalez2006dove}%
  \BibitemOpen
  \bibfield  {author} {\bibinfo {author} {\bibfnamefont {N.}~\bibnamefont
  {Gonz{\'a}lez}}, \bibinfo {author} {\bibfnamefont {G.}~\bibnamefont
  {Molina-Terriza}}, \ and\ \bibinfo {author} {\bibfnamefont {J.~P.}\
  \bibnamefont {Torres}},\ }\href@noop {} {\bibfield  {journal} {\bibinfo
  {journal} {Optics express}\ }\textbf {\bibinfo {volume} {14}},\ \bibinfo
  {pages} {9093} (\bibinfo {year} {2006})}\BibitemShut {NoStop}%
\end{thebibliography}%

\end{document}